# Laser-Cooling of Liquid Water by the Ar–Xe Laser Radiation

I.V. Kholin, D.A. Zayarnyi[*]

P.N. Lebedev Physical Institute of the Russian Academy of Sciences,

Leninskiy prospekt 53, Moscow, 119991, Russian Federation

**Abstract**

An effect of laser-cooling of water was observed for the first time with a temperature decrease $\Delta T = -2.2$ K after irradiation of liquid water surface by a powerful Ar–Xe pulse laser with a pulse energy of about 1 J and wavelength $\lambda = 1.73$, 2.63 and 2.65 μm. The discovered effect can apparently be ascribed to the optical excitation of vibrational states of $H_2O$ molecules followed by an endothermic consolidation of chemically active excited molecules into a quasi-stable cluster-like structure. The measured time dependences of the cooling effect show that a typical life time of the new state of water amounts to hours. It has also been shown that the life time of the excited vibrational molecular states due to a radiation trapping effect can be estimated to at least hundreds of seconds.

## 1. Introduction

The Ar–Xe high pressure lasers developed within the last few decades (see [1–4] and references in review [1]) are ranked among the most effective lasers for different kinds of applications in the near IR spectral range. Ar–Xe lasers demonstrate a high specific energy output (up to ~10 J/l [1, 3] with an efficiency exceeding 5% [1, 2]), a nearly diffraction-limited divergence of the output radiation (25–30 μrad [1–4]) and a convenient set of generated wavelengths [1, 2] in the near IR spectral range. Due to these merits the Ar-Xe laser system has a great potential for optical location, data communication and atmosphere monitoring.

This work started as a part of an investigation aimed at experimental study of the passage through the terrestrial atmosphere of powerful radiation with wavelengths 1.73, 2.63 and 2.65 μm. The 1.73 μm wavelength meets one of the most transparent Earth atmospheric windows. The two others, on the contrary, hit an intensive absorption band of the water vapor. Of great interest for us was also a process of interaction between the laser radiation and the liquid water which is a subject of the current article. Water is known to be one of the mostly widespread liquids on Earth. At the same time it possesses a set of unusual properties which distinguish it from other common liquids: anomalously high thermal capacity, cluster structure, *etc*. (for details see review [5] and references therein).

---

[*] Corresponding author email: dzayarny@sci.lebedev.ru

## 2. Experimental setup

The experiments were performed using an e-beam laser facility "Tandem". The gain medium of the Ar–Xe laser was excited by an electron beam with electron energy of about 250 keV. The cross-section of the electron beam was equal to $5\times100$ cm$^2$. The electron current pulse had a bell-shaped profile with approximately 2.5 μs duration at the base level. The current density amounted to 1.5 A/cm$^2$. The laser chamber (pos. *1* in Fig. 1) made of stainless steel had an active volume $5\times5\times100$ cm. The e-beam was injected into the laser chamber through a 20 μm-thick titanium foil. The BaF$_2$ chamber windows were mounted at the Brewster angle. An unstable telescopic resonator had a magnification ratio $M = 2$. A laser mixture was used with a component ratio Ar:Xe = 200:1 at a 2.5 atm pressure. The lasing spectrum consisted of three laser lines with $\lambda$ = 1.73, 2.63 and 2.65 μm and relative intensities $k_{1.73}$:$k_{2.63+2.65} \approx 8$:2. The total energy of a laser pulse was ~1 J. A more detailed description of the "Tandem" setup can be found elsewhere (see, *e.g.* [6]).

## 3. Experiments

*Temperature measurements using a thermocouple.* During the investigation of Ar–Xe laser radiation absorption distilled deionized water was placed in a flat glass cavity 15 mm in height and 100 mm in diameter (pos. *6* in Fig. 1, a thermocouple is not shown in the figure). The laser beam was focused by mirror *5* with a focal length $F = 15$ cm approximately normal to the water surface. The focusing point was inside the liquid so that the laser spot on the surface, which was slightly stretched due to spherical aberrations of the mirror, had a size of about 4–5 mm. This geometry prevented effects such as significant thermal flows due to laser breakdown on the water surface, water evaporation, splashing and stirring by the laser pulse. To stabilize the temperature regime we placed the optical scheme as a whole inside a closed wood box and kept it at constant temperature for about 12 hours before starting measurements.

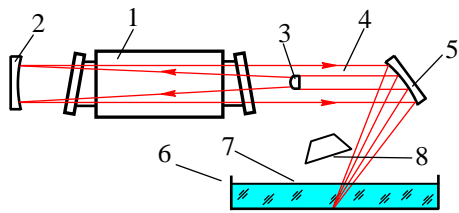

**Fig. 1.** Optical scheme of the experiment:

1 — laser chamber; 2 — concave mirror of the unstable resonator; 3 — convex mirror of the unstable resonator; 4 — laser beam; 5 — focusing mirror; 6 — experimental cavity; 7 — water surface; 8 — sensing surface of calorimeter.

An analysis of absorption spectra of water [7] shows that the Ar–Xe laser lines in use are located inside a wide range of absorption which corresponds to allowed vibrational transitions from the ground energy (000) level of H$_2$O molecules to the low stretching (100) and (001) states. Typical values of the radiation absorption factor for these wavelengths are 10–100 cm$^{-1}$. So the light is to be completely absorbed within a thin surface layer: about 1 mm of thickness for $\lambda = 1.73$ μm and ~0.1 mm thick for $\lambda = 2.63$ and 2.65 μm. It is significant that the experiments geometry secures against the physical contact between the water excited by the laser radiation and the cavity walls.

In the experiments we measured the water temperature inside the absorbing region as a function of time. The measurements were performed using a calibrated thermocouple with a thermojunc-tion size less then 0.05 mm which was placed about 0.1 –0.2 mm below the water surface. The thermal e.m.f. was recorded using a high-sensitivity dc current and voltage amplifier R341 and a PC-based oscilloscope FOSC-52A. In testing and calibrating the apparatus we measured temperature time dependences of vacuum oil VM-5C after laser irradiation (Fig. 2). From the oscilloscope trace in Fig. 2 one can see a fast heating of oil with a $\Delta T = 5.9$ K temperature increase occurring after the laser impact. A subsequent cooling-down of oil follows an exponential law with a characteristic time $\tau_o = 6.2$ s. The parameters $\Delta T$ and $\tau_o$ are in a good agreement with the estimates based on the measured laser pulse energy, the absorption factor, thermal capacity and thermal conductivity of vacuum oil known from specification of VM-5C oil.

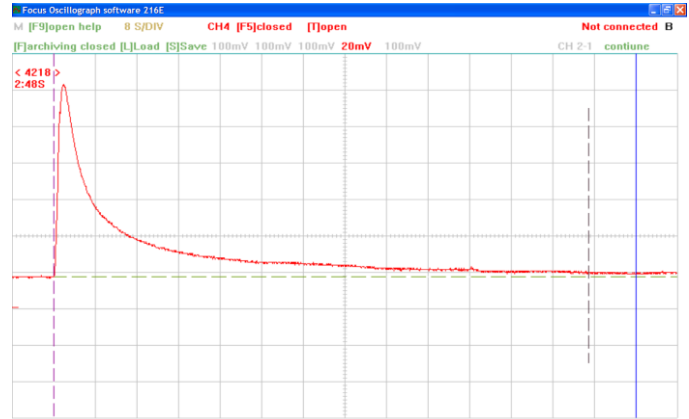

**Fig. 2.** Behavior of temperature change $\Delta T$ after laser irradiation of vacuum oil VM-5C (thermocouple measurements). The vertical scale is 1.13 K/div; horizontal scale is 8 s/div.

Temperature measurements in water showed quite a surprising result. Instead of expectable heating of the liquid and subsequent cooling down to the initial temperature we observed (as far as we know the first time) an effect of water **cooling** under the action of laser irradiation. The oscilloscope trace in Fig. 3 demonstrates that a small temperature increase which occurs at the leading edge of the laser pulse (due to a direct heating of the thermocouple by the laser radiation with $\lambda = 1.73$ μm) gives place to a temperature **decrease** to about 0.1 K below the initial level.

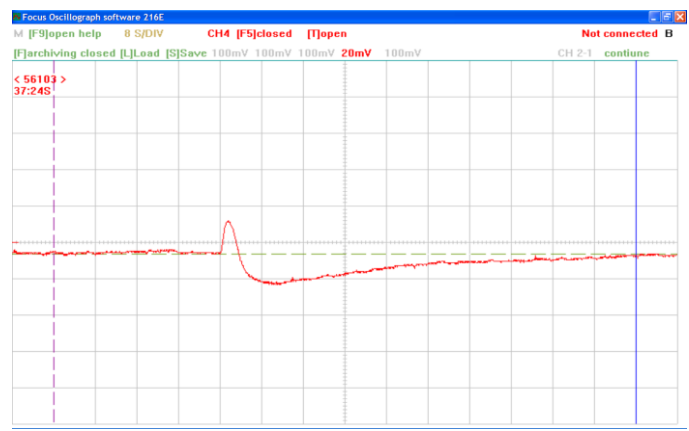

**Fig. 3.** Behavior of temperature change $\Delta T$ after laser irradiation of distilled deionized water (thermocouple measurements). The vertical scale is 0.11 K/div; horizontal scale is 8 s/div.

Such an anomalous behavior of the temperature-on-time dependence $\Delta T(t)$ can only be ascribed to interaction effects between the excited molecules before they relax to the ground state and the excitation energy is transferred into heat. As the collisions of the excited molecules with the thermocouple surface can strongly affect the relaxation processes, we continued our study of laser-cooling of water with non-contact methods using a high-sensitive calorimeter to measure the thermal radiation from the water surface.

*Non-contact temperature measurements.* During the non-contact measurements we replaced the thermocouple in the surface region of water by a high-sensitivity thermocouple calorimeter with a 2.5-cm-diameter sensing region which was placed in air near to the irradiated surface (see Fig. 1). To do this, the angle of incidence of the focused beam at the water surface was slightly increased. Such position of the sensor with respect to the laser spot on the surface enabled maximum amplitude of the recorded signal. Like in the case of thermocouple measurements, calibration experiments were performed using vacuum oil instead of water. The comparison of the signal measured by the calorimeter with the pulse obtained with the thermocouple allowed us to calibrate the measuring circuit.

In the case of laser irradiation of water (see Fig. 4) a fast heating and subsequent cooling of the calorimeter sensing plate were observed, as a consequence of a direct effect of the short laser pulse reflected from the water surface. Duration of this "forerunner" peak was determined by the response time of the calorimeter. Next, there is seen a process of long-term water cooling down to $\Delta T = 1.27$ K. The cooling effect duration was equal to $\tau_v \sim 200$ s. On reaching the minimum the water temperature started to rise slowly and returned back to the initial value at $\tau_w \sim 1600$ s.

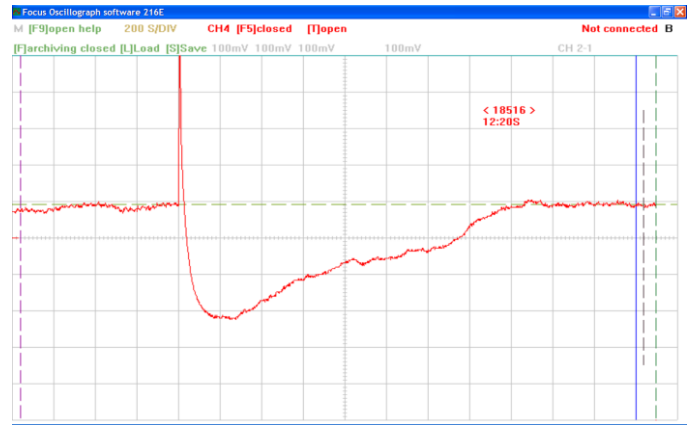

**Fig. 4.** Behavior of temperature change $\Delta T$ after laser irradiation of distilled deionized water (calorimeter measurements). The vertical scale is 0.41 K/div; horizontal scale is 200 s/div.

Efficient cooling of water occurs mostly due to the laser radiation with $\lambda = 2.63$ and $2.65$ μm. Setting into the laser beam a K8 optical glass filter, which absorbs the radiation with those wavelengths but transmits the radiation with $\lambda = 1.73$ μm, resulted in a multifold decrease of the cooling effect.

*Saturation of laser-cooling effect.* Fig. 5 shows a typical oscilloscope trace recorded after a multiple exposure of water to the laser pulses. In this experiment, seven laser shots were made with about 250 s time interval between the shots. The maximum cooling effect was observed after the first laser pulse. The effect decreased with each subsequent pulse and reached a saturation to the last shot. Finally a maximum cooling effect in these experiments $\Delta T = 2.2$ K and a maximum life time of the cooled water $\tau_w \sim 6000$ s were achieved.

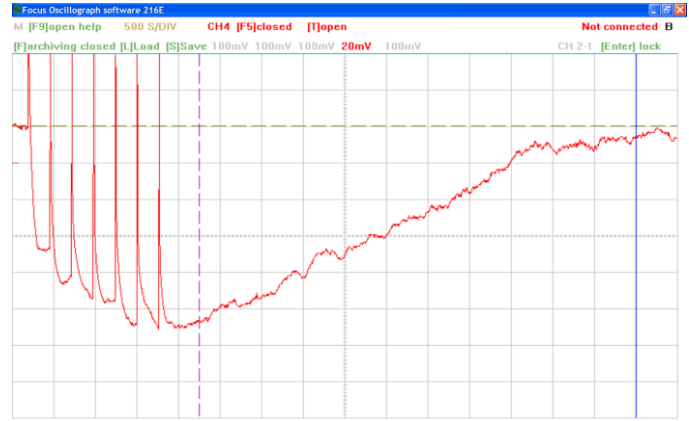

**Fig. 5.** Behavior of temperature change Δ*T* after irradiation of distilled deionized water by a series of laser pulses (calorimeter measurements). The vertical scale is 0.41 K/div; horizontal scale is 500 s/div.

## 4. Conclusion

The performed experiments allowed us to make following conclusions.

1. The experiments were carried out using distilled deionized water, and so the possibility of laser radiation absorption by foreign particles was totally prevented. Thus, the laser energy can only be spent on excitation by the allowed transitions to the vibrational states of $H_2O$ molecules. In our experiments there is no sign of water warming up within at least $\tau_v \sim 200$ s after the laser pulse termination. This indicates that the relaxation of the vibrational states of water to the ground energy level with the excitation energy being transferred to the translation degrees of freedom is a very slow process (in particular because of the radiation trapping effect on the allowed vibrational transitions).

2. According to the current knowledge, common unexcited water has a very complicated structure and consists of a conglomerate of short-lived molecular clusters which arise due to relatively weak hydrogen bonds [5]. The excited water molecules should possess an enhanced chemical activity compared to the molecules in the ground state. For this reason we relate the observed effect of laser-cooling in a thin film of vibrationally excited water to collisional endothermic processes accompanied by creation of stronger longer-lived hydrogen bonds between the $H_2O$ molecules. This leads to a consolidation of molecules into a quasi-stable cluster-like structure and creation of a new metastable water state with a formation time $\tau_v \sim 200$ s and a characteristic lifetime as long as a few hours.

3. An unusually large warming time of the cluster-like surface layer of the cooled water to the initial room temperature indicates absence of convection and a low rate of heat transfer from the main bulk of water due to the thermal conductivity process. The first fact unambiguously indicates that the specific weight of water is not being increased during the laser-cooling. At the same time we have not had sufficient experimental data to analyze the thermal conductivity processes at the moment, as we didn't measure the thickness of the cooled layer in our experiments.

4. The saturation of the laser-cooling effect indicates that virtually all molecules of the irradiated surface layer of water become involved in the creation of the cluster-like state after several laser pulses. This fact evidences that an effect of water bleaching under the laser irradiation at wavelengths $\lambda = 2.63$ and $2.65$ μm is possible. The existence of such a bleaching me-chanism can be beneficial for using the Ar–Xe laser radiation in monitoring the Earth atmosphere.

## Acknowledgements

The authors thank D.I. Kholin and N.N. Ustinovskii for useful discussions.